\documentclass[12pt]{iopart}
\usepackage{iopams,amsfonts,amssymb,amsthm}
\expandafter\let\csname equation*\endcsname\relax		
\expandafter\let\csname endequation*\endcsname\relax	
\usepackage{amsmath}

\usepackage[pdftex]{graphicx}       
\graphicspath{{figs/}}
\usepackage{mathrsfs}	

\DeclareMathOperator{\var}{var}

\usepackage{hyperref} 
\hypersetup{colorlinks,citecolor=blue,filecolor=blue,linkcolor=blue,urlcolor=blue}

\begin{document}

\title{Lattice polymers near a permeable interface}


\author{C J Bradly$^1$, N R Beaton$^1$ and A L Owczarek}
\address{
	$^1$School of Mathematics and Statistics, University of Melbourne, Victoria 3010, Australia}
\ead{
	\href{mailto:chris.bradly@unimelb.edu.au}{chris.bradly@unimelb.edu.au},
	\href{mailto:nrbeaton@unimelb.edu.au}{nrbeaton@unimelb.edu.au},
	\href{mailto:aleks.owczarek@gmail.com}{aleks.owczarek@gmail.com}
	}

\vspace{10pt}
\begin{indented}
\item[]\today
\end{indented}

\begin{abstract}
	We study the localisation of lattice polymer models near a permeable interface in two dimensions.
	Localisation can arise due to an interaction between the polymer and the interface, and can be altered by a preference for the bulk solvent on one side or by the application of a force to manipulate the polymer.
	Different combinations of these three effects give slightly different statistical mechanical behaviours.
	The canonical lattice model of polymers is the self-avoiding walk which we mainly study with Monte Carlo simulation to calculate the phase diagram and critical phenomena.
	For comparison, a solvable directed walk version is also defined and the phase diagrams are compared for each case.
	We find broad agreement between the two models, and most minor differences can be understood as due to the different entropic contributions.
	In the limit where the bulk solvent on one side is overwhelmingly preferred we see how the localisation transition transforms to the adsorption transition; the permeable interface becomes effectively an impermeable surface.
\end{abstract}



\section{Introduction}
\label{sec:Intro}

The statistical mechanics of lattice polymers near a surface is a long standing area of study.
If the surface is impermeable then the monomers may interact with the surface and a transition to an adsorbed phase occurs at a critical temperature \cite{Hammersley1982}.
On the other hand, the surface may be permeable, in which case we call it an interface and instead of adsorption there may be localisation to the interface.
Permeable interfaces in polymer systems model the interface between two immiscible solvents \cite{Iliev2004}.
The role of polymers at liquid-liquid interfaces arises in real world scenarios such as flocculation of pollutants in waste water treatment \cite{Lamanna2023}.

The canonical lattice model of polymers is the self-avoiding walk (SAW), so named for its restriction to not revisit lattice sites, which corresponds to the excluded-volume effect of real polymers.
However, the SAW model is not integrable and while some rigorous results can be derived, it is common to employ numerical methods, especially Monte Carlo simulation.
On the other hand, lattice polymer models based on directed or partially directed walks are useful since these models are often solvable, but at the cost of introducing a preferred direction that reduces the symmetry of the system.

In addition to the polymer interacting directly with the interface or bulk solvent another possible feature of lattice polymer models is the application of a force to deform or manipulate the chain, motivated by atomic force microscopy (AFM) \cite{Haupt1999,Zhang2003,Bemis1999}.
This can be modelled as the polymer compressed by a piston \cite{Guffond1997} but the force can be applied in multiple ways \cite{Orlandini2016}.
Here we consider the simplest case of one end tethered and the other pulled by a force perpendicular to the surface.
In lattice polymer models this induces a transition between pulling and pushing phases.
As the magnitude of the force increases the walks become ballistic, in the sense that the height of the pulled endpoint is $O(n)$ in the length of the walk $n$.
For self-avoiding walks the location of this transition is at zero force \cite{Beaton2015} and its critical properties are known \cite{Bradly2023}.
When applied to a polymer near a permeable interface the force can pull the polymer from one bulk phase to another across the interface \cite{Orlandini2004}.

To address these scenarios we consider models of lattice polymers with three different interactions.
Namely, an energy can be assigned to vertices in the thin interface, or different energies can be assigned to vertices on either side of the interface, or a work can be assigned to the height of the free endpoint pulled by a force.
In \sref{sec:Models} we introduce the directed walk and self-avoiding walk versions with some general remarks about how they are solved or simulated, respectively.
Then we consider three cases of the general model by omitting one of the three interactions.
Namely, in \sref{sec:PulledBisolvent} the {\em pulled--bisolvent} case has a force applied to the endpoint and variable bulk solvent below the interface but no interaction with the interface.
In \sref{sec:PulledLocalised} the {\em pulled--localised} case has the endpoint force and interaction with the interface, but neutral bulk on either side. 
In \sref{sec:LocalisedBisolvent} the {\em localised--bisolvent} case has interaction with the interface and variable bulk solvent below the interface but omits the endpoint force. 
In each case we compute the phase diagrams to compare the directed and self-avoiding walk models.
Finally, we summarise our findings in \sref{sec:Conc} and discuss the localisation transition. 

\section{General models}
\label{sec:Models}

The models we consider can be defined as special cases of a general model.
For each case of competing interactions, we will consider two versions, using a directed walk model and a self-avoiding walk model.
For walks of $n$ steps starting in the interface and ending at or above the interface, we also count the number $v$ of vertices below the interface, the number $m$ of vertices in the interface, and height $h$ of the endpoint above the interface.
The number of such directed walks is denoted $d_n(v,m,h)$ and the number of self-avoiding walks is denoted $s_n(v,m,h)$.
Conjugate to these parameters are weights: $z$ is conjugate to the length $n$, $b$ is conjugate to the number of bulk vertices below the interface $v$, $w$ is conjugate to the number of vertices in the interface $m$ and $y$ is conjugate to the endpoint height $h$.
The models we consider as special cases of this general model are obtained by omitting one of the parameters $v,m,h$, equivalent to setting its corresponding weight to 1.
Namely, the {\em pulled--bisolvent} model counts $h$ and $v$ but omits the number of vertices in the interface $m$; 
the {\em pulled--localised} model counts $h$ and $m$ but omits the number of bulk vertices below the interface $v$; 
and the {\em localised--bisolvent} model counts $m$ and $v$ but omits the endpoint height $h$. 

The directed walk version of the full model with all interactions has not been considered before to our knowledge.
However, all the special cases have been and no new insights are gained by considering all interactions simultaneously.
Furthermore, Monte Carlo simulations of SAWs cannot consider all interactions simultaneously.
This is because, for a lattice model with $p$ additional parameters besides length, our methods generate a $(p+1)$-dimensional histogram, with $O(n)$ size in each dimension. 
We can consider $p=2$ of the three additional parameters at a time and achieve useful lengths but the full SAW model is beyond our computational capability.
In order to utilise the directed walk model as a useful point of comparison we consider the special cases mentioned above.

\subsection{Directed walks}
\label{sec:GeneralDirected}

\begin{figure}
	\centering
	\includegraphics[width=0.75\textwidth]{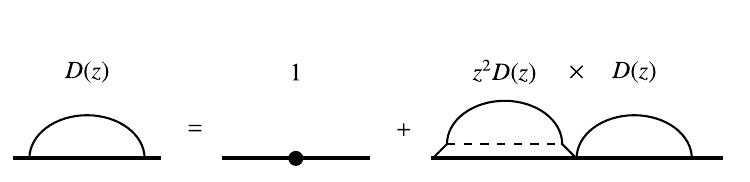}
	\caption{Factorisation of Dyck paths.}
	\label{fig:DyckFactors}
\end{figure}

For the directed walk model, $d_n$ counts the number of generalised Dyck paths which are paths in the square lattice that start at the origin and can step `up' in direction $(1,1)$ or `down' in direction $(1,-1)$.
The interface is the positive $x$-axis.
We define a generating function for such walks 
\begin{equation}
	Q(b,w,y,z) = \sum_{nvmh} d_n(v,m,h) \, z^n b^v w^m y^h.
	\label{eq:GFGeneral}
\end{equation}
One might also consider the number of vertices above the interface with some weight $a$, but here we consider the region above the interface to be neutral so $w$ and $b$ are defined relative to the bulk above the interface.
In particular, this means that $b$ indicates a variable preference to the bulk on either side of the interface and is not strictly modelling solvent quality which typically induces a collapse transition if the solvent quality is poor.

The basis of the directed walk models are Dyck paths which have the restriction that they cannot step below the interface and must terminate on the interface.
Recall that the generating function $D(z) = \sum_{n} d_n z^n$ of an ordinary Dyck path may be computed by a factorisation argument illustrated in \fref{fig:DyckFactors}. 
It corresponds to characterising any non-trivial Dyck path as composed of an initial sub-path that does not visit the interface, possibly followed by a further Dyck path.
Hence
\begin{align}
	D(z) &= 1 + z^2 D(z)^2 \nonumber \\
		&= \frac{1 - \sqrt{1 - 4z^2}}{2z^2}.
	\label{eq:GFDyck}
\end{align}
Extensions to this factorisation method will be applied to all directed walk models we consider in order to compute their generating functions.
The generating function $D(z)$ in \eref{eq:GFDyck} has a non-trivial singularity at $z = 1/2$, which is relevant to all models we consider here.
Other cases will contain additional relevant singularities that correspond to additional phases of the model.

\subsection{Self-avoiding walks}
\label{sec:GeneralSAW}

The self-avoiding walk model is also defined on the square lattice with walks starting at the origin.
Subsequent steps can be to any adjacent vertex so unlike the directed walk model a SAW can take steps within the interface, which is the entire $x$-axis.
For the SAW model it is useful to define a partition function
\begin{equation}
	Z_n(b,w,y) = \sum_{vmh} s_n(v,m,h) \, b^v w^m y^h,
	\label{eq:PartitionGeneral}
\end{equation}
which is related to a generating function via
\begin{equation}
	S(b,w,y,z) = \sum_n z^n Z_n(b,w,y).
\end{equation}
The parameters $v,m,h$ and weights $b,w,y$ have the same meaning as the directed walk model.

For the self-avoiding walk we use the flatPERM algorithm \cite{Prellberg2004,Campbell2020} to approximately enumerate $s_n(v,m,h)$.
This algorithm samples walks by starting at the origin and growing the endpoint using a pruning and enrichment strategy and measuring the parameters $v,m,h$.
The main output is a flat histogram $W_n$ that approximates $s_n$, for all lengths $n$ up to a preset maximum length.
For models that measure two of the three parameters $v,m,h$ the maximum length is limited since the histogram is three-dimensional; we simulate up to length $n = 128$ to obtain the phase diagram on a broad scale.
For some special cases we simulate up to length $n = 1024$ by only considering a two-dimensional histogram, counting length $n$ and vertices below the interface $b$.
In these cases we also fix the weighting $w$ of vertices in the interface for very little additional computational cost.
For all simulations we initiate 10 independent instances of flatPERM, each instance uses a parallel implementation with 8 threads, and each thread runs $10^5$ iterations.
This obtains a total of $\approx 2 \times 10^{11}$ samples at maximum length $n = 128$ for the first set of simulations or $\approx 6 \times 10^{10}$ samples at maximum length $n = 1024$ for the additional special cases.
The parallel implementation speeds up the initial equilibration phase \cite{Campbell2020} and incurs insignificant overhead since the parameter space is large enough for threads to avoid significant race conditions.

With the approximations to $s_n$ obtained from the simulations, the phase diagram of the self-avoiding walk model is determined by singular behaviour in its free energy
\begin{equation}
	f(b,w,y) = \lim_{n \to \infty} f_n(b,w,y), \quad f_n(b,w,y) = \frac{1}{n} \log Z_n(b,w,y).
\end{equation}
While some rigorous arguments can be made about the free energy of self-avoiding walk models, we mainly examine the singular behaviour by calculating order parameters $\langle q \rangle/n$, where $q \in \{v,h,m\}$.
For simulations that only have one such microcanonical variable (apart from length $n$) we also compute variances $\mathrm{var}(q) = \left( \langle q^2 \rangle - \langle q \rangle^2 \right)/n$.
For simulations with two of $v,h,m$ we calculate the largest eigenvalue of the covariance matrix
\begin{equation}
	H(\alpha, \beta) = \begin{pmatrix}
		\frac{\partial^2 f}{\partial \alpha^2} & \frac{\partial^2 f}{\partial \alpha \partial \beta} \\
		\frac{\partial^2 f}{\partial \beta \partial \alpha} &\frac{\partial^2 f}{\partial \beta^2}
	\end{pmatrix},
	\label{eq:Hessian}
\end{equation}
where $\alpha,\beta$ are the weights conjugate to the two microcanonical parameters chosen from $v,h,m$.
These quantities are calculated from the simulation data as weighted sums of moments
\begin{equation}
	\langle q^i \rangle (\alpha, \beta, \ldots) = \frac{1}{Z} \sum_{x,y,\ldots} q^i W_n(p,q,\ldots) \alpha^q \beta^p\ldots
	\label{eq:SimulationMoment}
\end{equation}
where $W_n$ is the approximation to $s_n$ produced by the flatPERM simulation.

\section{Pulled endpoint and bisolvent}
\label{sec:PulledBisolvent}

\subsection{Directed walks}
\label{sec:PulledBisolventDirected}

The first case we consider is the pulled--bisolvent case without interaction at the interface.
The directed walk model of this case was studied in \cite{Orlandini2004} and we summarise the relevant results here.
Non-trivial walks in this case are factorised into a bilateral Dyck path that can visit either side of the interface and ends at the interface, and then possibly a tail, which is a path from a vertex in the interface to a vertex of height $h$ conjugate to $y$, with all vertices (except the first) strictly above the interface.
The generating function of a bilateral Dyck path is determined by further factorising into a Dyck path which is the part of the walk before the first return to the interface that is strictly on one side of the interface, and then possibly the remainder of the walk which is also a bilateral path.
Hence the generating function is
\begin{align}
	B(b,z) &= \left(1 +  z^2 \left[ D(z) + b  D(b z) \right] \right) B(b,z) \nonumber \\
		&= \frac{1}{1 - z^2\left[ D(z) + b  D(b z) \right]},
	\label{eq:GFBilateralNoInterface}
\end{align}
where $D(z)$ is the generating function of an ordinary Dyck path, \eref{eq:GFDyck}.
The generating function for tails is
\begin{align}
	T(y, z) &= \frac{y z  D(z)}{1 - y z  D(z)} \nonumber \\
		&= \frac{2z}{2z - y \left(1 - \sqrt{1 - 4z^2} \right)}.
	\label{eq:GFTails}
\end{align}
Then the generating function for paths that can visit either side of the interface with endpoint at some height $h$ conjugate to $y$ is
\begin{align}
	G(b,y,z) &= \sum_{nbh} d_n(v,h) \, b^v y^h z^n, \nonumber \\
		&= \left[1 + B(b,z) \right] \left[1 + T(y, z) \right],
	\label{eq:GFPulledBisolvent}
\end{align}
This generating function has two singularities due to the bulk terms in $B$ and a third singularity from \eref{eq:GFTails}:
\begin{equation}
	z_1 = \frac{1}{2}, \quad z_2 = \frac{1}{2b}, \quad
	z_3 = \frac{y}{(y^2 + 1)},
	\label{eq:SingularitiesPulledBisolvent}
\end{equation}
corresponding to delocalised-above, delocalised-below and ballistic (above the interface) phases, respectively.
Where these singularities coincide determines the phase boundaries, and these are summarised as
\begin{equation}
\begin{tabular}{p{0.57\textwidth}ll}
	delocalised-above--ballistic (continuous): 				& $y_c = 1$, & $\quad b < 1$ \\
	delocalised-above--delocalised-below (first-order): 	& $b_c = 1$, & $\quad y < 1$ \\
	ballistic--delocalised-below (first-order): 	& $b_c = \frac{y^2+1}{2y}$, & $\quad y > 1$
\end{tabular}
\label{eq:BoundariesPulledBisolvent}
\end{equation}

\begin{figure}[t!]
	\centering
	\includegraphics[width=\textwidth]{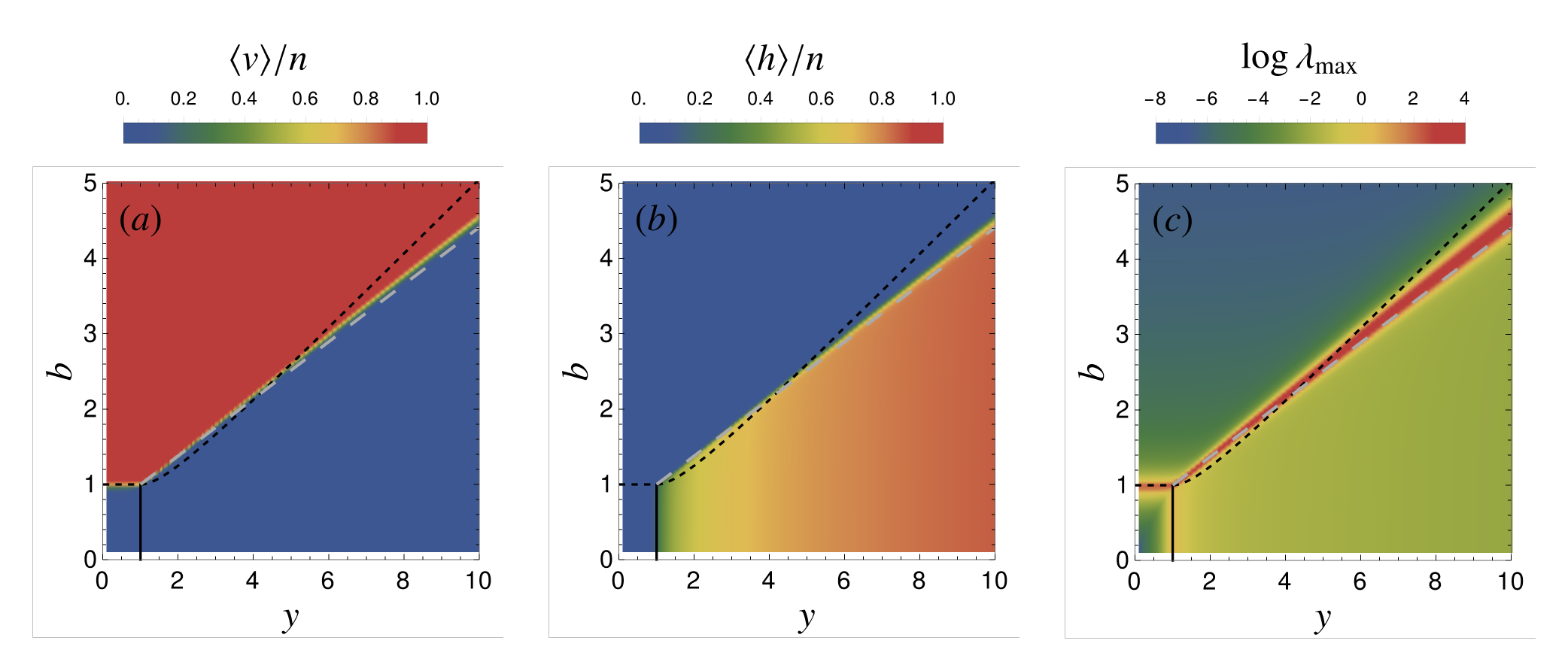}
	\caption{Phase diagram of walks with variable solvent quality below the interface and pulled from the endpoint.
	Order parameters (a) fraction below the interface $\langle v \rangle/n$ and (b) fractional endpoint height $\langle h \rangle /n$, as well as (c) the (logarithm of the) largest eigenvalue of the covariance matrix.
	Density plots are calculated from flatPERM simulation of the SAW model at length $n = 128$ and black lines are phase boundaries from the directed walk model \eref{eq:BoundariesPulledBisolvent}.}
	\label{fig:PulledBisolvent}
\end{figure}

\subsection{Self-avoiding walks}
\label{sec:PulledBisolventSAW}

For the self-avoiding walk model of the pulled--bisolvent case the partition function is
\begin{equation}
	Z_n(b, y) = \sum_{vh} s_n(v, h) \, b^v y^h,
	\label{eq:PartitionBisolvent}
\end{equation}
where the endpoint height of samples is restricted within the simulation to $h = \max(0, x_{2,n})$ so that $0 \leq h \leq n_\text{max}$.
Otherwise, values of the weight $y < 1$ would correspond to walks pulled ballistically below the interface.

We use flatPERM to estimate $s_n(v, h)$ and compute thermodynamic quantities.
In \fref{fig:PulledBisolvent} we plot order parameters (a) fraction below the interface $\langle v \rangle/n$ and (b) fractional endpoint height $\langle h \rangle /n$, as well as (c) the largest eigenvalue of the covariance matrix.
The phase boundaries of the directed walk model in \eref{eq:BoundariesPulledBisolvent} are overlaid for comparison.
The delocalised-below phase exists for both models in the region $y < 1$, $b < 1$ where both order parameters are close to zero.
If $b \leq 1$ is fixed but $y$ is varied, a continuous transition occurs for both models at $y_c = 1$ to the ballistic phase.
Similarly, for fixed $y \leq 1$ but $b$ is varied there is a first order transition to the delocalised-below phase at $b_c = 1$.

The ballistic--delocalised-below transition is also first-order in the SAW model but the location of the phase boundary is slightly different.
We know that the free energy of a ballistic polymer (in either model) has asymptotic form $f_\text{ball}(y) \sim \log y$ for large $y$.
For large $b$ walks are effectively confined to a half space below the interface, so each vertex has weight $b\mu$.
For SAWs, the delocalised-below phase thus has asymptotic free energy $f_\text{below}(b) \sim \log \mu b$ and so the boundary with the ballistic phase is at $b = y/\mu$ where $\mu \approx 2.63815853$ \cite{Clisby2013}.
A line with this slope is also plotted in \fref{fig:PulledBisolvent} and closely matches the boundary for the SAW model.
The boundary of the directed walk model is steeper since the growth constant for Dyck paths is exactly 2. This can also be seen from the expression for the ballistic--delocalised-below boundary in \eref{eq:BoundariesPulledBisolvent} which is asymptotically $b \sim y/2$ for large $y$.



\section{Pulled endpoint and interacting interface}
\label{sec:PulledLocalised}

The full model in Ref.~\cite{Orlandini2004} also allowed a separate weight $a$ for vertices visited above the interface.
In this case, localisation occurs when $a < 1$ and $b < 1$ and the interface is energetically favourable compared to the bulk on either side.
The generating function has an additional singularity in this case and if there is no pulling then there is a continuous transition when 
\begin{equation}
	b_c = \frac{a\left( 2a - 1 \right)}{2a^2 - 2a + 1},
	\label{eq:BoundaryABLocalisation}
\end{equation}
whereas if there is pulling there is a first-order transition at
\begin{equation}
	y_c = \frac{a^2b - a^2 -2ab + a + b}{a\sqrt{a-1}\sqrt{b-1}\sqrt{a + b - ab}}.
	\label{eq:BoundaryABLocalisationPulled}
\end{equation}

Instead of an additional weight to the bulk above the interface, our model has an explicit weight for vertices in the interface.
This is equivalent to the above with the rescaling $a \to 1$ above the interface, $1 \to 1/a = w$ in the interface and $b \to b/a$ below the interface.
In this section, we now consider the case where the bulk is neutral on both sides of the interface and only consider weight $w$ for vertices in the interface, corresponding to an interaction with a permeable surface, as well as the pulling force by weighting the endpoint with $y$.

\subsection{Directed walks}
\label{sec:PulledLocalisedDirected}

Directed walks with pulled endpoint and interaction with the interface are a bilateral walk that interacts with the interface, possibly followed by a tail with endpoint weight $y$.
The latter we considered above, see \eref{eq:GFTails}, and the former are bilateral Dyck paths with weight $w$ for vertices in the interface, but no bulk weighting.
These are either the trivial walk, or a Dyck path strictly on one side of the interface or the other, possibly followed by another bilateral path. 
The generating function for directed walks localised at a permeable interface is \cite{Rensburg2015}
\begin{align}
	I(w,z) &= 1 + 2 w z^2 D(1,z) I(w,z) \nonumber \\
			&= \frac{w}{1 - w \left( 1 - \sqrt{1 - 4z^2} \right)}.
	\label{eq:GFBilateralNoBulk}
\end{align}
Note that the starting vertex is fixed in the interface so the trivial walk does not contribute a weight $w$.

Combining the tail contribution and the bilateral walks we have for pulled and localised directed walks
\begin{align}
	D(w,y,z) &= \sum_{nmh} d_n(m, h) \, w^m y^h z^n \nonumber \\
		&= I(w,z) \left[ 1 + T(y,z) \right] \nonumber \\
		&= \frac{1}{1 - w \left( 1 - \sqrt{1 - 4z^2} \right)} \frac{2z}{2z - y \left(1 - \sqrt{1 - 4z^2} \right)}.
	\label{eq:GFPulledLocalised}
\end{align}
This generating function has three singular points
\begin{equation}
	z_1 = \frac{1}{2}, \quad
	z_2 = \frac{\sqrt{2w - 1}}{2w}, \quad
	z_3 = \frac{y}{y^2 + 1},
	\label{eq:SingularitiesPulledLocalised}
\end{equation}
corresponding to the delocalised, localised and pulled phases.
Note that in this case there is no orientation due to the bulk so walks in the delocalised phase will have, on average, half their vertices on either side of the interface.
The boundaries between these phases are
\begin{equation}
\begin{tabular}{p{0.4\textwidth}ll}
	delocalised--ballistic (continuous): 		& $y_c = 1$, & $\quad w \leq 1$ \\
	delocalised--localised (continuous):	& $w_c = 1$, & $\quad y \leq 1$ \\
	ballistic--localised (first-order): 	& $w_c = \frac{1}{2} \left(y^2 + 1\right)$, & $\quad y > 1$
\end{tabular}
\label{eq:BoundariesPulledLocalised}
\end{equation}

Note that this is a slight variation on adsorbing Dyck paths which are confined to the upper half-space \cite{Rechnitzer2004}.
In that case the factor of 2 is omitted in \eref{eq:GFBilateralNoBulk}, leading to an adsorption critical point at $w = 2$, whereas in our case the localisation critical point is at $w = 1$.

\subsection{Self-avoiding walks}
\label{sec:PulledLocalisedSAW}

The SAW model of pulled and localised walks has partition function
\begin{equation}
	Z_n(w, y) = \sum_{mh} s_n(m, h) \, w^m y^h,
	\label{eq:PartitionPulledLocalised}
\end{equation}
where the endpoint height is defined as $h = |x_{2,n}|$.
This is in contrast to the previous model where the different bulk solvent below the interface defines an orientation so that samples with endpoint below the interface have different physical meaning.
Such samples can be counted in this model since we now consider the `height' as the perpendicular distance from the interface.
While this choice avoids negative powers of $y$ in \eref{eq:PartitionPulledLocalised},  it is somewhat artificial in that the region $y < 1$ can not be interpreted as a downward force applied to the endpoint since that would push the endpoint through the permeable surface.
If samples above and below are treated as distinct then the phase diagram is symmetric about $\log y = 0$, but this scenario is less comparable to the other cases we consider.

\begin{figure}[t!]
	\centering
	\includegraphics[width=\textwidth]{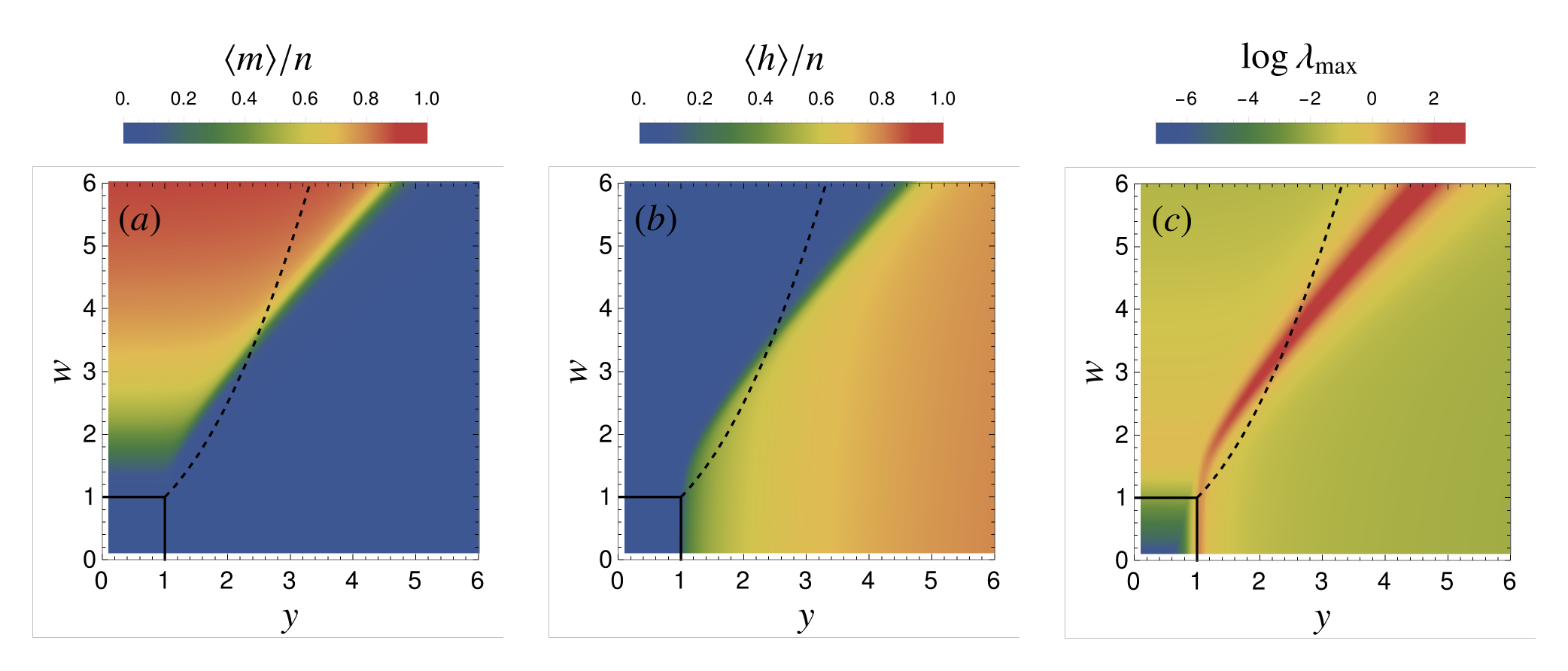}
	\caption{
	Phase diagram of walks pulled from a permeable interacting interface.
	Order parameters (a) fraction localised in the interface $\langle m \rangle/n$ and (b) fractional endpoint height $\langle h \rangle /n$, as well as (c) the (logarithm of the) largest eigenvalue of the covariance matrix.
	Density plots are calculated from flatPERM simulation of the SAW model at length $n = 128$ and black lines are phase boundaries from the directed walk model \eref{eq:BoundariesPulledLocalised}.
	}
	\label{fig:PulledLocalised}
\end{figure}

In \fref{fig:PulledLocalised} we plot thermodynamic quantities calculated from flatPERM simulations up to length $n = 128$.
Namely, order parameters (a) the fraction in the interface $\langle m \rangle/n$ and (b) fractional endpoint height $\langle h \rangle /n$, as well as (c) the largest eigenvalue of the covariance matrix.
The phase boundaries of the directed walk model in \eref{eq:BoundariesPulledLocalised} are overlaid for comparison.
The phases of these models are similar to the pulled--bisolvent case:
the localised phase exists for both models in the region $y < 1$, $w < 1$ where both order parameters are close to zero.
Then the ballistic phase exists for large $y$ at fixed $w$ where the localised fraction of the walk is zero, $\langle m \rangle / n \approx 0$ and the fractional endpoint height $\langle h \rangle/n \to 1$ as $y \to \infty$.
Similarly, the localised phase is where, for fixed $y$ and large $w$ the fractional endpoint height is zero, $\langle h \rangle / n \approx 0$, and the localised fraction increases $\langle m \rangle/n \to 1$ as $w \to \infty$.
The delocalised--ballistic transition at $y_c = 1$ and also the delocalised--localised transition are continuous for SAWs and directed walks.
The delocalised--localised transition is also continuous in both models.
The location of this boundary is clearly at $w = 1$ for the directed walk model, but its location in the SAW model is less clear from this data although it is expected to be in the same place \cite{Vrbova1998a}.
Even at $y = 1$ there are significant finite-size effects which make the delocalised--localised transition unclear.

The major difference between the SAW and directed walk models is the location of the ballistic--localised transition.
The boundary in the directed walk model shown in \eref{eq:BoundariesPulledLocalised} is a simple quadratic whereas the SAW boundary is asymptotically linear.
In the SAW case, walks in both phases have almost identical configurations; either a SAW is almost entirely perpendicular to the interface in the ballistic phase or it is parallel to the interface in the localised phase.
Hence there is no entropic contribution to the free energy when $y$ and $w$ are large.
By contrast, the directed walk in the localised phase (large $w$) still has an entropic contribution because only every second vertex can be in the surface; the other vertices can still be above or below.

\section{Interacting interface and bisolvent}
\label{sec:LocalisedBisolvent}

Next we omit the pulling, so $y = 1$, but the model is similar in that the competition between weighting the interface and weighting the bulk below acts in the same direction as a pulling force.

\subsection{Directed walks}
\label{sec:LocalisedBisolventDirected}

While directed walk models involving interactions with the interface and different bulk weights have been separately considered, to our knowledge their combination has not.
Using the same techniques as before, the generating function for this case is
\begin{align}
	B(b,w,z) &= \sum_{nvm} d_n(v,m) \, b^v w^m z^n, \nonumber \\
		&= 1 + w z^2 \left[ D(z) + b  D(b z) \right] B(b, w, z)\\
		&= \frac{2bw}{2b - w \left(1 - \sqrt{1 - 4z^2} \right) - b w \left( 1 - \sqrt{1 - 4b^2 z^2} \right)}.
	\label{eq:GFLocalisedBisolvent}
\end{align}
That is, the non-trivial walks are factorised into an initial segment that is either strictly above (the $D(z)$ term) or strictly below (the $b D(bz)$ term) the interface, possibly followed by a further non-trivial segment.
This generating function has two singularities due to the bulk on each side of the interface and a third pole singularity:
\begin{equation}
	z_1 = \frac{1}{2}, \quad z_2 = \frac{1}{2b}, \quad
	z_3 = \frac{\sqrt{w - 1} \sqrt{w - b} \sqrt{w + b w - b}}{w \left( w + b w - 2b \right)},
	\label{eq:SingularitiesLocalisedBisolvent}
\end{equation}
corresponding to the delocalised-above, delocalised-below and localised phases.
If $w > b \geq 1$ then the third singularity exists, and in particular, for $b = 1$ $z_3$ reduces to the second singularity in \eref{eq:SingularitiesPulledLocalised}. 
Otherwise the first two singularities in \eref{eq:SingularitiesPulledBisolvent} resolve to a phase boundary at $b = 1$ for $w < 1$, corresponding to a first-order transition between phases where the walks are delocalised either above or below the interface.
The phase boundaries may be summarised as
\begin{equation}
\begin{tabular}{p{0.5\textwidth}ll}
	deloc.-above--deloc.-below (first-order): 	& $b_c = 1$, & $\quad w < 1$ \\
	localised--delocalised-above (continuous): 			& $b_c = \frac{- w^2 + 2w}{w^2 - 2w + 2}$, & $\quad w > 1$ \\
	localised--delocalised-below (continuous): 	& $b_c = \frac{w^2 + w \left( \sqrt{w^2 + 4w - 4} - 2\right)}{4w - 4}$, & $\quad w > 1$
\end{tabular}
\label{eq:BoundariesLocalisedBisolvent}
\end{equation}

The localised--delocalised-below boundary has a complicated expression but its main property is that it quickly resolves to an asymptotic form $b \sim w/2$ as $w$ increases.
The delocalised-above--localised boundary joins the tricritical point $(w, b) = (1, 1)$ with the point $(w, b) = (2,0)$.
The latter is interesting because in the limit $b \to 0$ the walks are expelled from below the interface and the interface becomes effectively impermeable.
Given that vertices above the interface are also unweighted, in this limit the model is equivalent to adsorption of directed walks, which has known critical point at $w = 2$ \cite{Rechnitzer2004}.
One final remark is that the delocalised-above--localised boundary has a gradient of $-1$ at the point $(w, b) = (2, 0)$, in contrast to the SAW model discussed below.

\subsection{Self-avoiding walks}
\label{sec:LocalisedBisolventSAW}

\begin{figure}[t!]
	\centering
	\includegraphics[width=\textwidth]{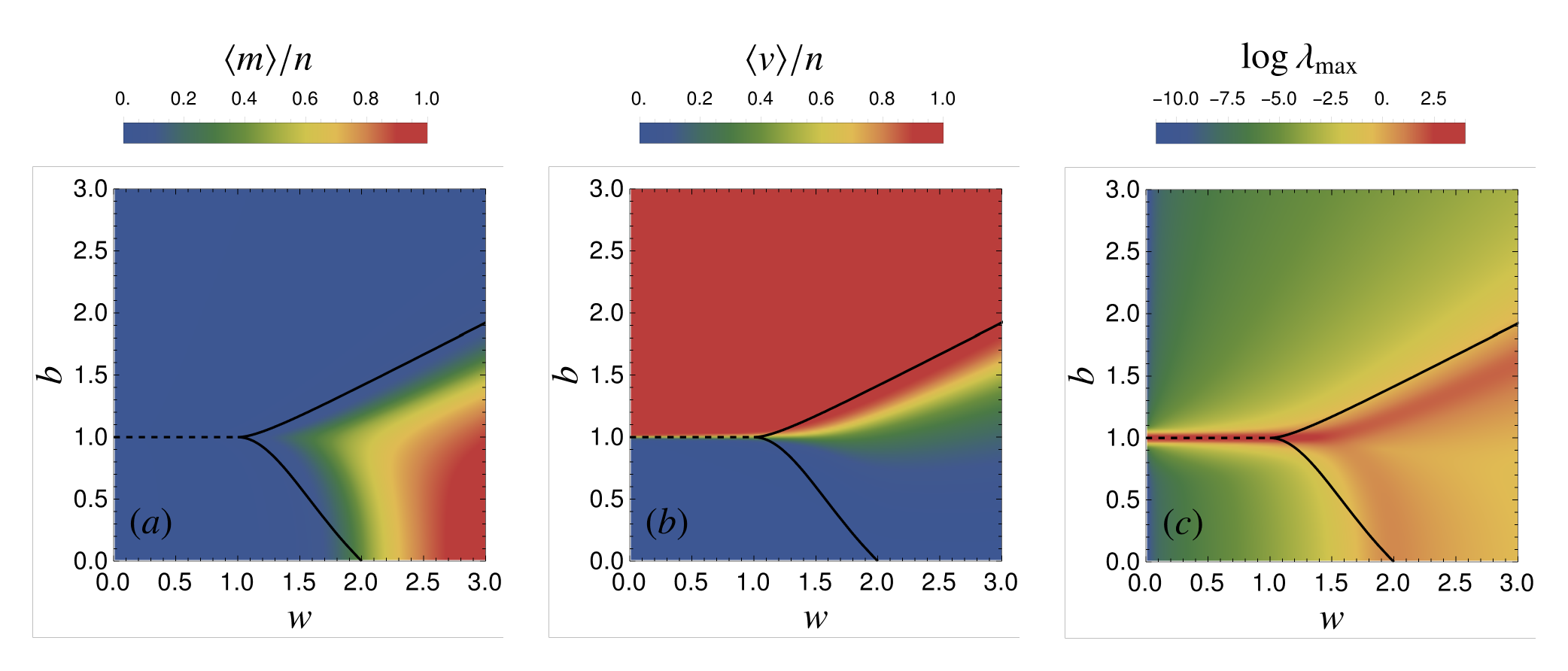}
	\includegraphics[width=0.4\textwidth]{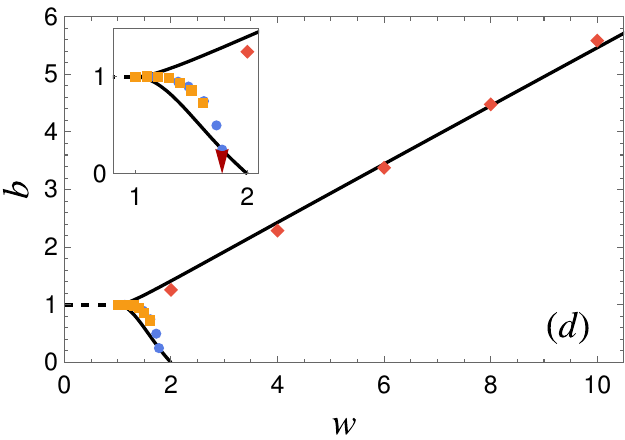}
		\caption{Phase diagram of walks interacting with a permeable interface and with variable solvent quality below the interface.
	Order parameters (a) fraction below the interface $\langle v \rangle/n$ and (b) fraction localised in the interface $\langle m \rangle/n$, as well as (c) the (logarithm of the) largest eigenvalue of the covariance matrix.
	Density plots are calculated from flatPERM simulation of the SAW model at length $n = 128$ and black lines are phase boundaries from the directed walk model \eref{eq:BoundariesPulledBisolvent}.
	(d) Locations of critical points from simulations up to $n = 1024$. Orange and red points are from simulations at fixed $w$, blue points are from simulations at fixed $b$. The inset shows the region where phase boundaries meet and the red arrow marks the location of the adsorption transition for an impermeable surface.}
	\label{fig:BisolventLocalised}
\end{figure}

The SAW model for walks interacting with a permeable surface and having bulk weight below the interface has partition function
\begin{equation}
	Z_n(b, w) = \sum_{vm} s_n(v, m) \, b^v w^m.
	\label{eq:PartitionBisolventLocalisation}
\end{equation}
Like the previous cases, simulation of this model indicates broad agreement with the directed walk model but there are some key differences.
The order parameters $\langle m \rangle/n$, $\langle v \rangle/n$ and covariance are shown in \fref{fig:BisolventLocalised} for SAWs of length $n = 128$, overlaid with the phase boundaries of the directed walk model in \eref{eq:BoundariesLocalisedBisolvent}.

For small $b$ and small $w$ there is a delocalised-above phase with both order parameters near 0.
For large $b$ the delocalised-below phase is characterised by $\langle v \rangle/n \approx 1$ and $\langle m \rangle/n \approx 0$.
Similarly, for large $w$ the localised phase shows $\langle m \rangle/n \to 1$ as $w$ increases, while $\langle v \rangle/n \to 0$ as $b$ decreases.
The boundaries between the phases show two key distinctions from the directed walk model, but the data for $n = 128$ is too imprecise for this purpose.
In order to resolve these differences, additional simulations were performed to longer lengths $n = 1024$ and with one of $b$ or $w$ fixed throughout the simulation, corresponding to horizontal and vertical slices through the phase diagram.
These additional simulations allow to estimate the location of critical points more accurately.
More precisely, the critical points are estimated by locating the peaks of the specific heats, calculated as the variance of the order parameters, $\var(m)/n$ and $\var(v)/n$, and then extrapolating to the limit of large $n$ using a finite-size scaling ansatz.
These additional points are shown in \fref{fig:BisolventLocalised}(d) along with the directed walk boundaries.
The blue points correspond to simulations at fixed $b$ (horizontal slices) and the orange and red points correspond to fixed $w$ (vertical slices), estimated in different intervals of $b$.
The inset is a zoom into the region where all three boundaries meet.

The first distinction from the directed walk model is that in the SAW model the delocalised-above phase pushes further into the localised phase as $w$ increases with $b$ just below 1, shown in the inset of \fref{fig:BisolventLocalised}.
The curvature of the boundary in this region is the reason we needed simulations corresponding to both vertical and horizontal slices in this region, and the agreement between blue and orange points is excellent.
At one end of the delocalised-above--localised boundary, near the tricritical point, the boundary line approaches the tricritical point horizontally as $w \to 1^+$, while at the other end, the boundary appears to approach $b = 0$ vertically.
For comparison, as discussed above, the localisation point of the directed walk model becomes the adsorption point at $w_\text{ads} = 2$, in the limit $b \to 0$ and has a finite gradient.
In the SAW model the adsorption point is $w_\text{ads} = 1.7743(35)$ \cite{Bradly2018}, which is marked on the inset of \fref{fig:BisolventLocalised} with a red arrow, and appears to be the limiting point of the boundary line.

The second distinction is in the asymptotic behaviour of the localised--delocalised-below boundary.
The boundary in the directed walk model has asymptotic form $b_c(w) \sim w/2$, by \eref{eq:BoundariesLocalisedBisolvent} for large $w$.
For the SAW model, the red diamonds in \fref{fig:BisolventLocalised}(d) are estimates of the location of the transition $b_c(w)$ from simulations at fixed $w = 2,4,6,8,10$.
These points suggest the relation $b_c(w) \sim 0.54\,w$ for the SAW model boundary.

\begin{figure}[t!]
	\centering
	\includegraphics[width=\textwidth]{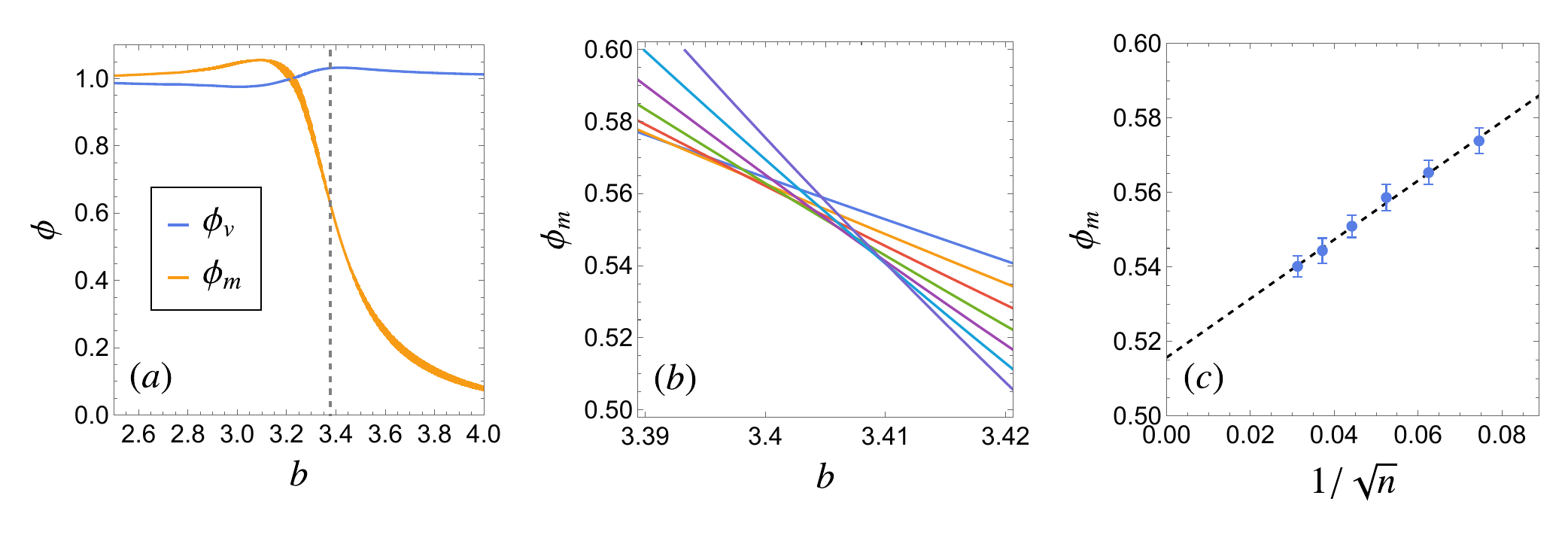}
	\caption{Critical properties of the localised--delocalised-below transition at fixed $w = 6$.
	(a) Exponents estimated from length scaling of order parameters. Dashed line marks the estimate of critical point from peaks of the specific heat.
	(b) Exponent $\phi_m$ estimated from \eref{eq:ExponentPhiRatio} for $n = 180,\ldots,1024$, error bars are consistent for all points in this range and are omitted for clarity.
	(c) Estimates of $\phi_m$ at critical point from intersections of curves in (b). Dotted line is a fit to extrapolate to $n \to \infty$.}
	\label{fig:ExponentsBisolventLocalised}
\end{figure}

In addition to the location of the boundary we can also use the larger $n$ data to probe the nature of the localised--delocalised-below transition.
The standard expectation is that the order parameters have finite-size scaling
\begin{equation}
	\frac{\langle m \rangle}{n} \sim n^{\phi_m - 1}, \quad \frac{\langle v \rangle}{n} \sim n^{\phi_v - 1},
	\label{eq:OrderParameterScaling}
\end{equation}
where the leading order exponents $\phi_m$ and $\phi_v$ should be universal.
We can estimate their values directly by fitting the order parameters using the data from simulations at fixed $w$.
The behaviour of the transition is independent of $w$ in the region $w \gtrsim 2$ so for example we show results for $w = 6$ only.
In \fref{fig:ExponentsBisolventLocalised}(a) we plot estimates for $\phi_m$ and $\phi_v$ across the localised--delocalised-below transition.
These curves are calculated by fitting the data to \eref{eq:OrderParameterScaling}, without corrections to scaling which in this case do not have a significant effect.
In both phases as well as at the critical point the data is consistent with $\phi_v = 1$ everywhere.
That is,  $\langle v \rangle/n \approx 0$ in the localised phase and  $\langle m \rangle/n \approx 1$ in the delocalised-below phase without any significant dependence on $n$.
On the other hand, $\phi_m = 1$ in the localised phase ($b < b_c$) where $\langle m \rangle/n > 0$ and $\phi_m = 0$ in the delocalised-below phase ($b > b_c$) where $\langle m \rangle/n \approx 0$.
The vertical dashed line is the location of the critical point $b_c(6) = 3.377(6)$ determined from extrapolating the location of peaks of $\var(m)/n$ as discussed above.
At this point the exponent has an intermediate value, namely $\phi_m = 0.631(4)$.
However, the value varies quickly near the critical point and is sensitive to inaccuracy in the value of $b_c$.

A better estimate of the exponent $\phi_m$ can be obtained directly by inverting \eref{eq:OrderParameterScaling} since we have data for different lengths $n$:
\begin{equation}
	\phi_m^{(n)} = 1 + \log_2 \frac{u_n}{u_{n/2}},
	\label{eq:ExponentPhiRatio}
\end{equation}
where $u_n = \langle m \rangle / n$.
We calculate $\phi_m^{(n)}$ as a function of $b$, shown in \fref{fig:ExponentsBisolventLocalised}(b). 
Intersections of these curves for different $n$ indicate the value of the exponent without preknowledge of $b_c$ by other means.
However, the values of $\phi_m^{(n)}$ at these intersection points can be extrapolated to $n \to \infty$, as shown in \fref{fig:ExponentsBisolventLocalised}.
By considering all points along the localised--delocalised-below boundary for $w \geq 2$ the average extrapolated value is $\phi_m = 0.512(3)$.
This is close to $\phi_m = 1/2$, and if we further assume the tricritical scaling relation $2 - \alpha = 1/\phi$, then this implies that the critical exponent $\alpha \approx 0$.
These critical exponents are the same as the adsorption transition in two dimensions where, as in that case, $\alpha = 0$ means that the peaks of the specific heat (which generally scale as $n^{\alpha\phi}$) are not suitable for locating the critical point with enough accuracy to also estimate $\phi$.
On the scale of the phase diagrams in \fref{fig:BisolventLocalised} this inaccuracy is not important for establishing the phase boundaries, but for a more thorough analysis of the critical properties we would need to employ additional estimates like for the adsorption transition \cite{Bradly2018}.

We have already discussed that as $b \to 0$ the localisation transition becomes the adsorption transition since the interface becomes an effectively impermeable interface.
The same is true for $b \to \infty$ except that the preferred side is now below the interface.
For SAWs confined to a half-space the free energy is the same as SAWs on the entire lattice, so an extra weight $b$ means that the free energy of weighted half-space walks is 
\begin{equation}
	f_\text{half}(b) = \log \left(\mu_d b \right)
	\label{eq:FreeEnergyHalfSpace}
\end{equation}
On the other hand, for walks confined to a $(d-1)$-dimensional interface with additional weight $w$ on each vertex
\begin{equation}
	f_\text{interface}(w) = \log \left(\mu_{d-1} w \right).
	\label{eq:FreeEnergyInterface}
\end{equation}
The free energy $f(w, b)$ of SAWs with both interaction at the interface and variable bulk solvent below the interface depends on the phase:
\begin{equation}
	f(b, w) 
	\begin{cases}
		= \log \left(\mu_d \right) & \text{$b$ small, $w < w_c(b)$},  \\
		= \log \left(\mu_d b \right) & \text{$b$ large, $w < w_c(b)$}, \\
		\sim \log \left(\mu_{d-1} w \right) & \text{$w \to \infty$, $w > w_c(b)$, $b$ fixed},
		\label{eq:FreeEnergyBisolventLocalised}
	\end{cases}
\end{equation}
corresponding to the delocalised-above, delocalised-below and localised phases, respectively.
The boundary $w_c(b)$ between phases depends on $b$ but this does not immediately indicate what $w_c$ should be.
In either delocalised phase, if $w$ is small, i.e. $w \sim O(1) < w_c$ the walk will also avoid the interface, but if $w > w_c$ then some positive fraction of the walk will be in the interface. 
In the latter case the value of $w_c$ is different whether coming from the delocalised-below or delocalised-above phase because of how we have defined the weights in our model.
More precisely, for the delocalised-above phase (small $b$) we have seen that $\lim_{b\to 0} w_c = w_\text{ads}$.
If the delocalised-below phase has the same behaviour then for large $b$ we must have also
$\lim_{b\to \infty} w_c/b = w_\text{ads}$.
In fact this is consistent with our data in that the slope of the localised--delocalised-below boundary in \fref{fig:BisolventLocalised} is close to $b_c \sim w/w_\text{ads} \approx 0.56w$.

If we also consider the weight of vertices above the interface we can map between the delocalised-above and -below phase diagrams with the transformation
\begin{equation}
	(a = 1, b, w) \mapsto (1/b, 1, w/b)
	\label{eq:RescalingTransformation}
\end{equation}
which interchanges which side has neutral bulk weighting.
These arguments also apply to the directed walk model; under the transformation \eref{eq:RescalingTransformation} the expressions for the localised--delocalised-below and localised--delocalised-above boundaries in \eref{eq:BoundariesLocalisedBisolvent} are symmetric.
This is shown in \fref{fig:SchematicPhases}(c) which illustrates the underlying symmetry.

\section{Conclusion}
\label{sec:Conc}

\begin{figure}[t!]
	\centering
	\includegraphics[width=\textwidth]{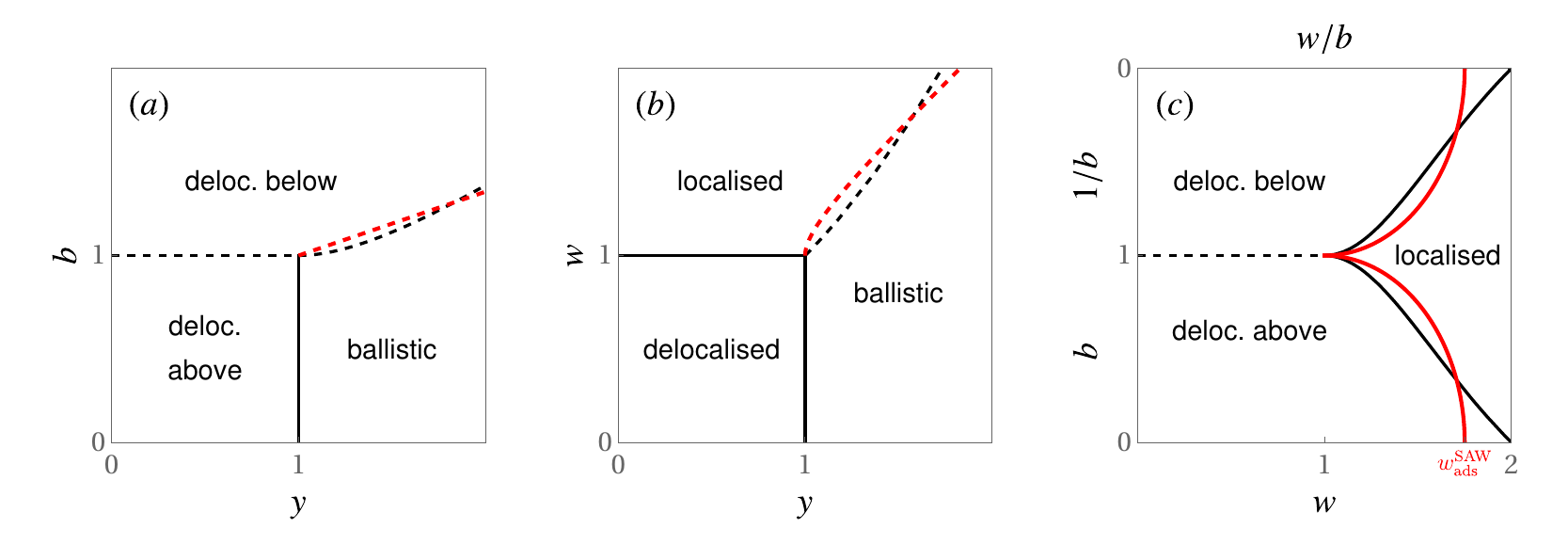}
	\caption{Schematic phase diagrams for the different cases (a) pulled--bisolvent, (b) pulled--localised and (c) localised--bisolvent.
	In the latter case the upper half has been transformed to indicate the symmetry in delocalised phases.
	Black lines are phase boundaries for the direct walk models, and red lines are where the SAW model is different. 
	Dashed lines indicate first-order transitions. }
	\label{fig:SchematicPhases}
\end{figure}

We have studied a number of models of lattice polymers that involve a permeable interface and thoroughly mapped their statistical mechanical properties.
Direct interactions with the interface, variable bulk solvent below the interface and the application of a manipulating force applied to the endpoint were all considered.
In the case of the directed walk model, all three interactions can be considered together, and for completeness the generating function for the full model defined in \eref{eq:GFGeneral} can now be determined as
\begin{equation}
	Q(b,w,y,z) = \left[ 1 + B(b,w,z) \right] \left[ 1 + T(y,z) \right],
	\label{eq:GFGeneralResolved}
\end{equation}
with $B(b,w,z)$ and $T(y,z)$ from \eref{eq:GFLocalisedBisolvent} and \eref{eq:GFTails}, respectively.
There are no additional singularities of the full model that aren't covered by \eref{eq:SingularitiesPulledLocalised} and \eref{eq:SingularitiesLocalisedBisolvent}.
However, in order for clarity and comparison to the SAW models, we studied two of the three interactions at a time.
Schematic phase diagrams for the different cases are summarised in \fref{fig:SchematicPhases}.
The phase diagrams of the two models have fairly good qualitative agreement in most cases.
The transitions between non-delocalised phases are generally continuous, unless a pulling force is applied, in which case the transition is first-order.
Minor differences in location of phase boundaries between the SAW and directed walk models are due to the different entropic contributions to the free energy of each model.
For the localised-bisolvent case careful analysis of the critical properties of the localised-delocalised transitions suggest a finite-size crossover exponent $\phi = 1/2$ in which case estimation of the location of the critical point is difficult without a more comprehensive study.
However, we do have enough accuracy to resolve the region of the localised--bisolvent phase diagram where all phase boundaries meet and in particular see how the localisation transition transforms to the adsorption transition in the limits $b \to 0$ and $b \to \infty$ where the permeable interface becomes effectively an impermeable surface.
As such the upper half of the phase diagram has been transformed via \eref{eq:RescalingTransformation} to indicate the symmetry between the delocalised-above and delocalised-below phases.

\begin{figure}[t!]
	\centering
	\includegraphics[width=\textwidth]{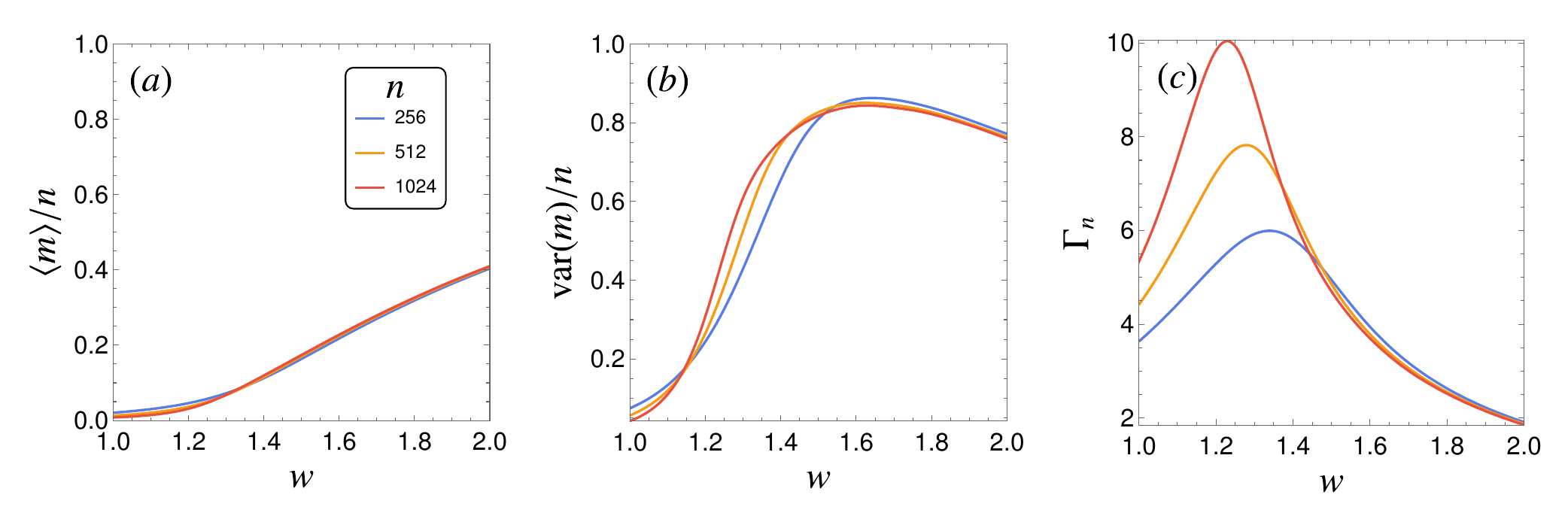}
	\caption{Thermodynamic quantities near the localisation transition for SAWs up to $n=1024$. (a) Localised fraction, (b) variance and (c) log-derivative or localised fraction.}
	\label{fig:LocalisationOnly}
\end{figure}

Without adulteration by other interactions (i.e. $y = 1$, $b = 1$), the localisation transition of lattice polymers is expected to be at $w_c = 1$ \cite{Vrbova1998a,Madras2017}.
In the pulled--localised model, the localisation transition for $y \leq 1$ appears to be at $w_c > 1$, possibly due to finite size effects.
In the localised--bisolvent model we were able to resolve the region near the tricritical point at $(w, b) = (1, 1)$, at least enough to distinguish the different phase boundaries.
One of the blue points in \fref{fig:BisolventLocalised} is for exactly $b = 1$ and closer inspection of this particular case indicate significant finite-size effects.
This particular point is showcased in \fref{fig:LocalisationOnly} where the location of peaks in $\var (m)/n$ or $\Gamma_n = \partial_w \log (\langle m \rangle /n)$ show very slow convergence to a transition at $w = 1$.

More generally, we may consider localisation of $d$-dimensional walks to a permeable $\delta d$-dimensional hypersurface.
Then, with $\Delta = d - \delta d$, it is expected that the critical point for SAWs is $w_c = 1$ if $\Delta = 1$ \cite{Madras2017}. 
Our preliminary Monte Carlo results for the SAW model in higher dimensions agree with this but show significant finite-size effects like those of the two-dimensional case seen in \fref{fig:LocalisationOnly}.
We also see the same behaviour for $\Delta = 2$.
That is, 3D walks interacting with a line ($\Delta = 2$) or plane ($\Delta = 1$), or 4D walks interacting with a 2D plane ($\Delta = 2$) are not inconsistent with $w_c = 1$ but display very large finite-size effects that make precise estimation of $w_c$ difficult.
One might expect the issue to be that flatPERM does not efficiently sample high-dimensional walks that visit a lower-dimensional subspace, but we have also found that simulations of 4D walks interacting with a line ($\Delta = 3$) clearly indicate $w_c > 1$ and the analysis is not inhibited by the kind of finite size effects seen in \fref{fig:LocalisationOnly}.

As for directed paths, there is an argument to suggest that $w_c > 1$ when $d > 3$ and $\delta d = 1$\footnote{Consider a projection of the directed walk to the $(d-1)$-dimensional hyperplane orthogonal to the interacting line, with the line mapping to the origin. Without interactions, the projected walk is essentially a simple random walk on some kind of lattice, which is recurrent if $d-1\leq 2$ but transient if $d-1>2$. Hence a weight $w_c>1$ is required to induce a positive density of returns to the origin.}, but for $2 < \delta d < d$ the value of $w_c$ may be sensitive to the orientation of the $\delta d$-dimensional hypersurface.

It may be the case that the localisation transition by itself can be better understood with significant dedication of computational power, which we did not employ in this work due to the range of cases considered. 
A thorough investigation of the SAW localisation transition and its finite-size scaling will be explored in future work.












\ack	
The authors acknowledge financial support from the Australian Research Council via its Discovery Projects scheme (DP230100674).
This research was supported by The University of Melbourne's Research Computing Services and the Petascale Campus Initiative.
Data generated for this study is available on request.

\section*{References}		
\bibliographystyle{plain}
\IfFileExists{../../../papers/bibtex/polymers_master.bib}
{\bibliography{../../../papers/bibtex/polymers_master.bib}}
{\bibliography{models.bib}}

\begin{thebibliography}{10}

\bibitem{Beaton2015}
Nicholas~R Beaton.
\newblock The critical pulling force for self-avoiding walks.
\newblock {\em J. Phys. A: Math. Theor.}, 48(16):16FT03, 2015.

\bibitem{Bemis1999}
Jason~E. Bemis, Boris~B. Akhremitchev, and Gilbert~C. Walker.
\newblock Single polymer chain elongation by atomic force microscopy.
\newblock {\em Langmuir}, 15(8):2799--2805, April 1999.

\bibitem{Bradly2023}
C.~J. Bradly and A.~L. Owczarek.
\newblock Critical behaviour of the extended-ballistic transition for pulled
  self-avoiding walks.
\newblock {\em Physica A}, 624:128978, 2023.

\bibitem{Bradly2018}
C.~J. Bradly, A.~L. Owczarek, and T.~Prellberg.
\newblock Universality of crossover scaling for the adsorption transition of
  lattice polymers.
\newblock {\em Phys. Rev. E}, 97:022503, Feb 2018.

\bibitem{Campbell2020}
S~Campbell and E~J {Janse van Rensburg}.
\newblock Parallel {PERM}.
\newblock {\em J. Phys. A: Math. Theor.}, 53(26):265005, jun 2020.

\bibitem{Clisby2013}
Nathan Clisby.
\newblock Calculation of the connective constant for self-avoiding walks via
  the pivot algorithm.
\newblock {\em J. Phys. A: Math. Theor.}, 46(24):245001, 2013.

\bibitem{Guffond1997}
M.~C. Guffond, D.~R.~M. Williams, and E.~M. Sevick.
\newblock End-tethered polymer chains under afm tips: Compression and escape in
  theta solvents.
\newblock {\em Langmuir}, 13(21):5691--5696, October 1997.

\bibitem{Hammersley1982}
J~M Hammersley, G~M Torrie, and S~G Whittington.
\newblock Self-avoiding walks interacting with a surface.
\newblock {\em J. Phys. A: Math. Gen.}, 15(2):539, 1982.

\bibitem{Haupt1999}
B.~J. Haupt, J.~Ennis, and E.~M. Sevick.
\newblock The detachment of a polymer chain from a weakly adsorbing surface
  using an {AFM} tip.
\newblock {\em Langmuir}, 15(11):3886--3892, 1999.

\bibitem{Iliev2004}
G.~Iliev, E.~Orlandini, and S.~G. Whittington.
\newblock Adsorption and localization of random copolymers subject to a force:
  The morita approximation.
\newblock {\em Eur. Phys. J. B}, 40(1):63--71, 2004.

\bibitem{Rensburg2015}
E~J {Janse van Rensburg}.
\newblock {\em The Statistical Mechanics of Interacting Walks, Polygons,
  Animals and Vesicles}.
\newblock Oxford University Press, 2015.

\bibitem{Lamanna2023}
Leonardo Lamanna, Gabriele Giacoia, Marco Friuli, Gabriella Leone, Nicola
  Carlucci, Fabrizio Russo, Alessandro Sannino, and Christian Demitri.
\newblock Oil-water emulsion flocculation through chitosan desolubilization
  driven by ph variation.
\newblock {\em ACS Omega}, 8(23):20708--20713, June 2023.

\bibitem{Madras2017}
Neal Madras.
\newblock Location of the adsorption transition for lattice polymers.
\newblock {\em J. Phys. A: Math. Theor.}, 50(6):064003, 2017.

\bibitem{Orlandini2004}
E~Orlandini and S~G Whittington.
\newblock Pulling a polymer at an interface: directed walk models.
\newblock {\em J. Phys. A: Math. Gen.}, 37(20):5305, may 2004.

\bibitem{Orlandini2016}
E~Orlandini and S~G Whittington.
\newblock Statistical mechanics of polymers subject to a force.
\newblock {\em J. Phys. A: Math. Theor.}, 49(34):343001, 2016.

\bibitem{Prellberg2004}
Thomas Prellberg and Jaros\l{}aw Krawczyk.
\newblock Flat histogram version of the pruned and enriched {Rosenbluth}
  method.
\newblock {\em Phys. Rev. Lett.}, 92:120602, Mar 2004.

\bibitem{Rechnitzer2004}
A.~Rechnitzer and E.J. {Janse van Rensburg}.
\newblock Exchange relations, dyck paths and copolymer adsorption.
\newblock {\em Discrete Appl. Math.}, 140(1):49--71, 2004.

\bibitem{Vrbova1998a}
Tereza Vrbová and Stuart~G Whittington.
\newblock Adsorption and collapse of self-avoiding walks at a defect plane.
\newblock {\em J. Phys. A}, 31(34):7031, aug 1998.

\bibitem{Zhang2003}
Wenke Zhang and Xi~Zhang.
\newblock Single molecule mechanochemistry of macromolecules.
\newblock {\em Prog Polym Sci}, 28(8):1271 -- 1295, 2003.

\end{thebibliography}

\end{document}